\documentclass[aps,prl,twocolumn,groupedaddress,showpacs]{revtex4}

\usepackage{graphicx}
\begin{document}

\title{Pairing with Unconventional Symmetry around BCS-BEC Crossover: Fermionic Atoms in 2D 
Optical Lattices to Correlated Electron Systems}


\author{Du\v san Vol\v cko}
\author{Khandker F. Quader}
\affiliation{Department of Physics, Kent State University, Kent, OH
44242}


\date{\today}

\begin{abstract}
We study superfluid properties of fermions on a 2D lattice using a finite-range 
pairing interaction derivable from an extended Hubbard model. We obtain signatures of 
unconventional pair-symmetry states, $d_{x^2-y^2}$ and extended-s ($s^*$), 
in the BCS-BEC crossover region. 
The fermion momentum distribution function, $v_k^2$, the ratio of the Bogoliubov coefficients,
$v_k/u_k$, and the Fourier transform of $v_k^2$ are among the properties that are
strikingly different for d- and $s^*$ symmetries
in the crossover region. 
Fermionic atoms in 2D optical lattices may provide a way to observe these signatures.
We discuss possible experimental ramifications of our results.

\end{abstract}


\maketitle

Attainment of boson and fermion condensates~\cite{bei} of ultracold neutral atoms
has presented an unprecedented opportunity to study properties of
quantum many-particle systems. {\it Fermionic} atoms in {\it optical lattices}~\cite{greiner02,kohl05}
constitute yet another intriguing set of systems.
While these are by themselves interesting to study, they may also provide a way to 
gain useful insight into properties of correlated electrons 
in solids.  Jaksch et al~\cite{jaksch98} suggested that atoms in optical lattices, confined 
to the lowest Bloch band, can be represented by the Hubbard model with 
hopping kinetic energy $t$ between neighboring sites, 
and on-site interaction $U$.
Hubbard model calculations~\cite{micnas,theory,kotliar88} predict that attractive-U
Hubbard model give rise to s-wave superconductivity, while the repulsive-U model results
in an antiferromagnetic or a d-wave superconducting phase depending on filling (number of
fermions per lattice site). Owing to the continuous tunability of model parameters such as,
density, hopping or interactions,  
optical lattices can serve as testing grounds for such models.
This has led, for example, to the suggestion~\cite{hofstetter02} that the
underlying physics of the high $T_c$ superconductors may be understood by 
studying these systems. 
Recent work~\cite{kohl05,duan05,multiband} have pointed out possible
role of additional Bloch bands and multi-band couplings 
in optical lattices. In solids this would correspond to having multiple orbitals and
near-neighbor interactions. Duan~\cite{duan05} has shown that on different sides of a 
broad Feshbach resonance, the effective Hamiltonian can be reduced to a t-J model, familiar
in correlated electron systems, wherein it has been suggested~\cite{kotliar88} that t-J model
can give rise to d-wave pairing.

Fermionic atoms subjected to positive and negative detuning using Feshbach resonance 
technique provide realizations of BEC-BCS crossover behavior.
It has been recently suggested~\cite{euro} that it should also 
be possible to study superfluid properties
of fermions in {\it optical lattices} around BEC-BCS crossover regime.
Starting with the seminal work of Eagles~\cite{eagles} and Leggett~\cite{leggett}, 
the BEC-BCS crossover problem received considerable theoretical 
attention~\cite{micnas,VQ,becbcshtccont,melo,becbcshtclat,derhertog99,andrenacci99} 
due to the possibility that high $T_c$ superconductors, possessing short coherence lengths,
could fall in the BEC-BCS crossover region. Several authors employed 
continuum models~\cite{micnas,VQ,becbcshtccont,melo,andrenacci99},
focussing mostly on conventional s-wave pair symmetry. Lattice models with on-site
or nearest-neighbor attractions have also been 
considered~\cite{micnas,VQ,becbcshtclat,derhertog99,andrenacci99}. 
More recent theory work~\cite{becbcscold} 
are in the context of cold fermions.

Motivated by these issues, {\it in this paper}, 
we study superfluid properties of fermions in a 2D square lattice
in the BEC-BCS crossover regime using a finite-range {\it pairing} interaction, 
obtainable from a multi-band {\it extended} Hubbard model. 
As representative cases of unconventional pair symmetry,
we consider two even-parity representations of the
cubic group, namely the $\ell =2$ $d_{x^2-y^2}$-wave, and the
$\ell=0$ extended s-wave ($s^*$). 
There has been work~\cite{derhertog99,andrenacci99,chen} employing similar pairing interaction;
however these have focussed on different systems and issues. We present several new results,
including specific signatures of superfluid states with 
{\it unconventional pairing gap}  symmetry as one goes between the BEC and BCS regimes.
This could provide a way to distinguish between different gap symmetry states
in systems that allow for tuning into the BEC-BCS crossover regime, such as fermionic atoms in
2D optical lattices, and possibly high $T_c$ cuprates.
One of our key results is
the remarkable behavior of the fermion distribution function, $v_k^2$,
(related to momentum distribution, $n_k$):
For the d-wave gap function,  $v_k^2$ changes {\it abruptly} from
having a peak at the Brilloiun zone (BZ) center (0,0)
to a vanishing central peak accompanied by a
redistribution of the weight around other parts of the
BZ ($(0,\pm \pi)$,$(\pm \pi, 0)$) as the system crosses from
the weak-coupling BCS to the strong-coupling BEC regime.
By contrast, $v_k^2$ changes smoothly in the $s^*$-wave case.
Similar signatures are also found in the ratio of Bogoliubov coefficients $v_k/u_k$,
related to the phase of the superfluid wavefunction. The Fourier transform of $v_k^2$
in real space exhibits a ``checkerboard'' type pattern that could have consequences for experiments.

The extended Hubbard model for two equal species population system on a 2D square lattice 
is given by:
\begin{eqnarray}
H&=&\sum_{<ij>\sigma}(-t c_{i\sigma}^+c_{j\sigma}+\rm{H.c.}) + U\sum_{i}n_{i\sigma}n_{i-\sigma}\nonumber\\
&-&V\sum_{<ij>\sigma\sigma^{\prime}}n_{i\sigma} n_{j\sigma^{\prime}} - \mu_{o}\sum_{i}n_{i},
\end{eqnarray}
where $t$ is the kinetic energy hopping, 
$\mu_{o}$ the unrenormalized chemical potential,
$U$ the on-site repulsion and $V$ the nearest-neighbor attraction. In the case of
cold fermions on a lattice, $V$ would be related to inter-band coupling.
$\sigma$ is the ``pseudo-spin'' index, that could refer to equally populated hyperfine states
in the case of optical lattices. 
At the mean-field level, the Hartree self-energy terms
renormalize $\mu_{o}$ such that $\mu = \mu_o + \mu_U(f) +\mu_V(f)$
where $\mu_U(f)$ and $\mu_V(f)$ are filling-dependent corrections
to $\mu$. We work with the
renormalized $\mu$ so as to properly deal with weak and strong couplings, and
take $\mu_{J_i}(f) = J_if$, where $J_i = U$, $- V$.
The filling $f=N/2M$, with $N$ the number of particles, $M$
the number of lattice sites, and the pseudo-spin degeneracy factor 2.
On Fourier transforming and retaining interactions between
particles with equal and opposite momentum, as in BCS theory,
the reduced {\it pairing Hamiltonian} assumes the form:
\begin{equation}
 H_{pair}=\sum_{k}(\epsilon_k-\mu) c^{+}_{k}c_k+\sum_{kk'}V_{kk'}c^{+}_{k'}c^{+}_{-k'}
c_{-k}c_{k}
\end{equation}
where in the tight-binding approximation,
$\epsilon_k=- 2t (\cos k_x+\cos k_y)$; $V_{kk'}=V_0 (\cos(k_x-k'_x)+\cos(k_y-k'_y))$, which
is {\it non-separable}. 
Using the standard BCS variational ansatz,
$|\Phi_{BCS}> = \prod_{\bf k} (u_{\bf k} + v_{\bf k}c_{\bf k}^{{\dag}}
c_{\bf - k}^{{\dag}})|0>$,
we obtain the $T=0$ gap equations
for the gap functions $\Delta_k^{d,s}=\Delta_o (f) (\cos k_x \pm \cos k_y)$
with $d_{x^2-y^2}$(-) and $s^*$(+) symmetries,
 
\begin{equation}
\frac{1}{V_o} =
\frac{1}{2 M} \sum_{k}^{\rm{BZ}}\frac{\cos k_x (\cos k_x \pm \cos k_y)}
{E_k^{d,s^*}},
\end{equation}
where
$E_k^{d,s^*} =
((\epsilon_k-\mu)^2+\Delta_{o}^2 (\cos k_x \pm \cos k_y)^2)^{1/2}$.
The Bogoliubov coefficients are given by,
\begin{equation}
|u_{\bf k}|^2; \;\; |v_{\bf k}|^2 = \frac{1}{2}
(1 \pm \frac{\epsilon_k -\mu}{E_k^{d,s^*}}).
\end{equation}
The ratio $v_{\bf k}/u_{\bf k} = - (E_k^{d,s^*} - (\epsilon_k - \mu))/
\Delta_k^{d,s^*}$. 
Following Leggett\cite{leggett}, we readjust $\mu$ for strong
attractions by supplementing the T=0 gap equation with the number equation:
\begin{equation}
N= \sum_{k}^{\rm{BZ}} (1-(\frac{\epsilon_{k}-\mu}{E_k^{d,s^*}});
\end{equation}
This determines the self-consistently readjusted $\mu$, which is no longer
fixed at the Fermi level, and makes the gap
equation applicable over the entire range of filling, thereby the BCS and BEC regimes. 
To allow for strong scattering, the sums are performed over the entire BZ.
The natural momentum cut-off afforded by the lattice avoids any possible
ultraviolet divergences.
 
\begin{figure}
      \centering
      \includegraphics[scale=0.85]{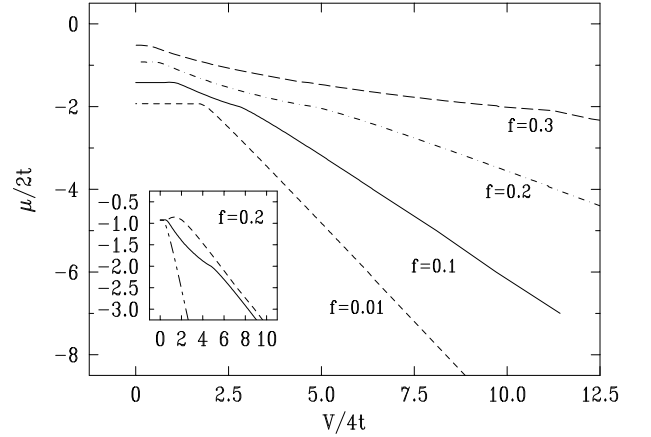}
      \caption{Chemical potential $\mu$ vs. coupling V  at different
fillings $f$  for the $d$-wave case.
BEC pairs appear where $\mu(V)$ crosses the $\mu/2t =-2$ line.
The \underline{inset} shows $\mu(V)$ for the $s$- (dash-short dashed line),
$s^*$ (dashed line), and d-wave (solid line) at
$f=0.2$.}
\label{fig:Fig.1}
 \end{figure}

Remarkable differences in  features stem
in an essential way from differences in gap symmetry.
The $d_{x^2-y^2}$ gap $\Delta_k^d$ vanishes
along the lines $\pm k_x = \pm k_y$
in the 2D BZ, i.e. at {\sl four} points on the Fermi surface(fs), the location
of which depends upon filling.
The $s^*$ gap $\Delta_k^{s^*}$
coincides with the tight-binding fs at exact 1/2-filling,
and is nodeless otherwise. Here, $\mu \le 0$,
with $\mu = -4t$ at the bottom of the band. Owing to particle-hole
symmetry, it is sufficient to consider $0 \leq  f \leq 1/2$.
Upon examination of the gap functions and Eqs. (3-5), the
following {\it distinctions} become apparent:
 
(a) For very low fillings
($f \rightarrow 0, \mu \rightarrow -4t$), a threshold coupling is required
for pairing in the d-wave case, while in the \underline{$s^*$ case}
$\Delta^{s^*} \rightarrow 0$ as $V \rightarrow 0$ due to a
{\sl weak singularity}
at $\mu =-4t$. On the other hand, at 1/2-filling,
due to a {\sl weak singularity} at $\mu = 0$
in the \underline{d-wave case}, $\Delta^d \rightarrow 0$ as $V \rightarrow 0$.
In the $s^*$ case such a singularity is not present and
as $\Delta^{s^*} \rightarrow 0$, $V/4t \rightarrow \pi^2/8$, i.e.
a minimum coupling is needed for pairing.
In contrast with $\Delta_o^{s^*}(V)$, $\Delta_o^{d}(V)$ changes
slope at $\mu = -4t$, and hence not smooth everywhere (though continuous).
 
(b) For small $k$, we have the following limiting behavior:
(i) $\epsilon_k < \mu(= -4t); \;\; |u_k| \rightarrow 1, |v_k| \rightarrow 0$;
this is the {\it strong-coupling} BEC limit. Here
the ratio $v_k/u_k \sim \Delta_k/2|\mu| \rightarrow (k_x^2 - k_y^2)/2|\mu|$,
i.e. {\it analytic}.
(ii) $\epsilon_k > \mu(= -4t); \;\; |u_k| \rightarrow 0, |v_k| \rightarrow 1$;
this is the {\it weak-coupling} BCS limit. Here
$v_k/u_k \rightarrow 1/(k_x - k_y)$, i.e. {\it non-analytic}.
(iii) $\epsilon_k = \mu(= -4t); \;\; |u_k| \ne 0, |v_k| \ne 0$,
when $E_k \rightarrow 0$.
Then $v_k/u_k  \sim (k_x - k_y)/(k_x + k_y)$, i.e. intermediate between
(i) and (ii).
It may be noted that for d-wave, the quasiparticle excitations
in the BCS limit  (ii) are ``gapless'' for some values
of $k$, while in the  BEC limit (i),
$E_{k} \ne 0$, even for gaps with nodes~\cite{Mohit}.
 
Self-consistent numerical solutions of Eqs.(3-5) bear out the above features in
detail, and also reveal a number of {\it other features}. We scale $\mu$, $V$, $\Delta$
by hopping parameter, $t$.
At a given filling $f$, both $\Delta_k^d$ and $\Delta_k^{s^*}$ increase
with increasing $V$. While for d-wave it is
easier to pair electrons at
higher fillings, this is not necessarily the case for
$s^*$-wave for the weaker couplings $V/4t \leq 1.5$ and
small gaps $\Delta^{s^*}/2t \leq 0.5$.

In Fig.1 we show $\mu(V)$ for different fillings f. At a fixed $f$,
in both the $d$- and $s^*$-wave cases, $\mu$ decreases with increasing
coupling $V$, changing less rapidly for progressively larger $f$.
However in the $s^*$ case, $\mu(V)$ exhibits a small ``bump'' for
weaker couplings $V/4t \leq 1.5$.
The drop in $\mu$ with increasing attraction is
significantly more rapid in the uniform
$s$-wave case; see inset in Fig. 1.
{\it Crossover to the BEC regime} here is signalled by
$\mu(V)$ going below the bottom of the band,
i.e. crossing the $\mu = - 4t$ line.
As Fig. 1 shows, for the d-wave case, this develops at both
low and high fillings at some minimum value
$V_b/4t$ of the coupling.
It is interesting to note that as
$f \rightarrow 0$, $V_b/4t \rightarrow 1.8$.
At exactly 1/2-filling this coupling tends to infinitely
large values.
For couplings $V > V_b$, the system is
conducive to BEC pairing;
for $V < V_b$, the system exhibits BCS-like features.

\begin{figure}
      \centering
      \includegraphics[scale=0.85]{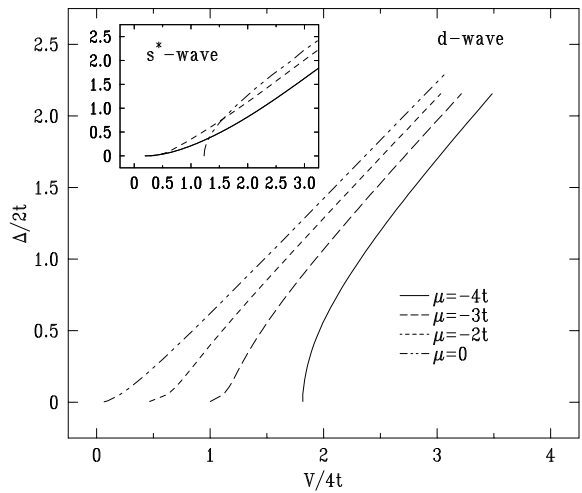}
      \caption{d-wave gap functions $\Delta/2t$ vs. nearest-neighbor coupling
$V/4t$ for different chemical potential $\mu$. \underline{Inset}: Results for the $s^*$ case.
$\mu = - 4t$ demarcates BEC and BCS regimes.}
\label{fig:Fig.2}
 \end{figure}
 
Fig. 2 shows the behavior of the d-wave gaps as a function of coupling
$V$ for different values of the chemical potential $\mu$.
The $\mu=-4t$ curve represents the locus of $V_b/4t$ for different
fillings (see Fig. 1), and demarcates BEC and BCS -pair regimes.
To the left is the $\mu>-4t$ region wherein finite gaps of the BCS or intermediate
BCS-BEC types exist. On a given
constant-$\mu$ curve it may not be possible to have solutions for
any arbitrary filling,
but only those that satisfy Eqs. (3) and (4) self-consistently.
The {\it inset} in Fig. 2 shows the corresponding $\Delta^{s^*}(V)$ curves
for the $s^*$ case. There are interesting differences with the d-wave results
in that the boundary ($\mu=-4t$) separating BEC/BCS regimes 
is not as clear-cut for the weaker couplings $V/4t \leq 1.5$ and
the smaller gaps $\Delta/2t \leq 0.5$, however the $\mu<-4t$ region
lies to the right of the $\mu=-4t$ curve as in the $d$-wave case.
 
\begin{figure}
     \centering
     \includegraphics[scale=0.40]{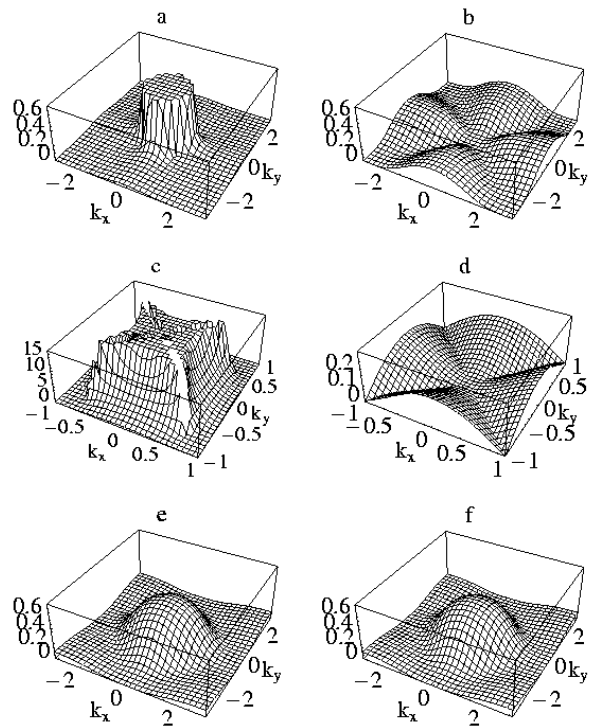}
     \caption{(a), (b): 3D plots of d-wave electron distribution
functions $v_k^2$ vs. $k_x-k_y$ at filling $f=0.1$, showing
abrupt ``jump'' in $v_k^2$. In
BCS regime (a), $\mu = - 3t$, $\Delta^d = 5.2t$, $ V = 18.7t$, and in
BE regime (b), $\mu = - 6t$, $\Delta^d = 0.6t$,
$ V = 5.2t$.
(c), (d): 3D plots of d-wave $v_k/u_k$ vs $k_x-k_y$ for the same parameters
as in ((a),(b)) respectively. In BCS regime (c)
it can be seen to be non-analytic; in BEC regime (d)
it is analytic. (e), (f): The same as in (a), (b), but for
$s^*$-wave; the behavior is smooth.}
\label{fig:Fig 3}
     \end{figure}

Differences in the gap symmetry manifest in a
striking manner in the momentum distribution function, $v_k^2$, and
the ratio $v_k/u_k$.
For d-wave, for a given filling, in the {\sl weak-coupling} BCS regime
($V < V_b(f)$, $\mu > -4t$), $v_k^2$ exhibits a peak centered
around the zone center (0,0),
that becomes progressively narrower with decreasing filling.
Then at the crossover point
at $V_b(f)$ ($\mu=-4t$), $v_k^2$ {\it abruptly}
goes to zero around (0,0), and shows a drastic redistribution in a different
region of the BZ, namely, along $(0, \pm \pi)$, and $(\pm \pi, 0)$.
The abruptness is evident from the ``jump''
in $v_k^2$ as the chemical potential
goes from just above the bottom of the band ($\mu > -4t$) to just
below ($\mu < -4t$), i.e. from BCS to BEC regime. A representative case is shown in Figs. 3a,3b.
In marked contrast, in the the $s^*$-wave case (Figs 3e,3f), the zone center peak
in $v_k^2$ decreases {\it smoothly} as one goes from the BCS regime
to the BEC regime; only a slight redistribution occurs at $(\pm \pi, \pm \pi)$.
We find this behavior to be replicated at all fillings f.
As observed above in the limiting cases, the numerical
calculations show (Fig 3c,3d) that for d-wave,
in the {\it weak-coupling} BCS regime, $v_k/u_k$ is {\it non-analytic}
at $\pm k_x = \pm k_y$; in the {\it strong-coupling} BEC regime,
$v_k/u_k$ is {\it analytic},
vanishing along the zone diagonals and peaking about
$(\pm \pi,0)$, $(0, \pm \pi)$.
In the $s^*$ case (not shown), $v_k/u_k$ is analytic in both regimes.
Similar behavior in $n_k$ has also been reported in other work~\cite{melo,VQ,derhertog99}.

\begin{figure}
     \centering
     \includegraphics[scale=0.30]{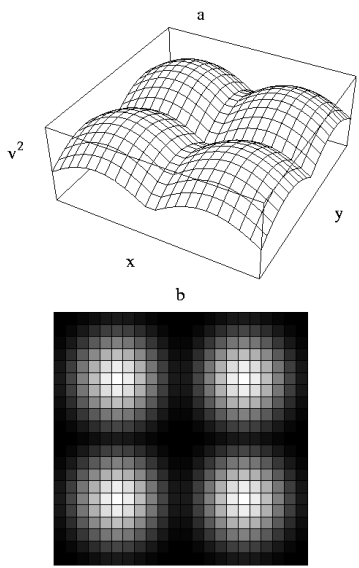}
     \caption{(a) Fourier transform $\rho_v(x,y)$ of a typical d-wave
electron distribution function, $v_k^2$. Here, filling $f=0.01$,
$\mu = - 4.2t$ (strong-coupling regime), gap $\Delta = .76t$. (b) Projection
of (a) to show contrast ratio of $\rho_v(x,y)$.}\label{fig:Fig 4}
     \end{figure}

Our findings suggest that experiments, that may
be able to directly or indirectly
probe $v_k^2$ or combinations of $u_k$ and $v_k$, could
reveal novel aspects of the paired states. For example,
it may be possible to decipher the OP symmetry (e.g. d- or $s^*$- wave)
by measuring $v_k^2$ as a function of
filling (especially at low-fillings),
and/or for different interaction strengths, both of which can be controlled in optical
lattices. 
At the BCS-BEC crossover, we expect the behavior
to be quite different depending
on whether the OP is d- or $s^*$ wave.
Also, in the case of d-wave pairs,
quantities sensitive to $v_k^2$ or to ($u_k, v_k$) should be very different
depending on whether the paired state is BEC or BCS like.
A possible probe may be ARPES. Information may also be obtained
from experiments that sample
the quasiparticle energy $E_k = (\Delta_k^2 + \epsilon_k^2)^{1/2}$ (related
to $u_k,v_k$), the quasiparticle density of states, or
coherence factors,   $u_kv_k +u_k'v_k'$.
Angle-dependent or transverse ultrasound attenuation\cite{coleman},
or quasiparticle tunneling at low fillings
are possible experiments.

The Fourier transform of $v_k^2(k_x,k_y)$, namely, $\rho_v(x,y)$ may provide 
yet another interesting way to test our results. In the {\it d-wave case}, in marked contrast 
with its behavior in the BCS regime,
$\rho_v(x,y)$ is {\it oscillatory} in the BEC regime, and exhibits an inhomogenous
``checkerboard-type'' pattern as shown in Fig 4(a,b).
For the chosen parameters of Fig. 3, the contrast ratio 
of the lowest density to the peak is roughly 50\%, being
most sensitive to the location of $\mu(V)$. The length scale
is of the order of fractions of lattice spacing.
$\rho_v(x,y)$ is fairly uniform in the {\it $s^*$ case} in both regimes.
Highly sensitive STM may be able to pick up such distinctions\cite{STM}. 

Much of the phenomena we have discussed
are away from exact 1/2-filling, and at
relatively strong coupling, where possible effects of spin density
wave (SDW) and charge density wave (CDW) instabilities are expected to
be suppressed.   Addition of a next-near-neighbor hopping
would also stabilize the paired state, as well as lower the minimum
near-neighbor interaction necessary for a bound-state;
we have checked this\cite{VQ}. We have not explored here the issues
of collective modes or phase separation\cite{phasesep}.
It may be interesting to
extend this work to, for example, finite-T, or to explore whether
the inhomogenous density that we find bear relationship to
the range/strength  of the interaction, or to possible phase separation.
 
We thank E. Abrahams, S. Davis, and H. Neuberger for discussions and comments.
The work is supported in part by ICAM.

\end{document}